\documentclass[useAMS,usenatbib,usegraphicx]{mn2e}
\usepackage{amsmath}
\newcommand{\sect}{\S\,}

\newcommand{\e}[1]{10$^{#1}$}

\newcommand{\kms}{~km\,s$^{-1}$}

\newcommand{\msun}{M$_{\odot}$}

\title[Compact object collisions]
{Going out with a bang: compact object collisions
resulting from supernovae in binary systems}

\author[Troja et al.]{E. Troja$^{1,2}$, G. A. Wynn$^{2}$, P. T. O'Brien$^{2}$,
S. Rosswog$^{3}$ \\
$^{1}$INAF - Istituto di Astrofisica Spaziale e Fisica Cosmica,
     Sezione di Palermo, via Ugo la Malfa 153, 90146 Palermo, Italy\\ 
$^{2}$Department of Physics and Astronomy, University of Leicester,
      Leicester, LE1~7RH, UK\\
$^{4}$School of Engineering and Science, Jacobs University Bremen, 
Campus Ring 1, 28759 Bremen, Germany}
		 
\begin{document}

\date{Accepted -- -- --. Received -- -- --.}

\pagerange{\pageref{firstpage}--\pageref{lastpage}} \pubyear{2009}

\maketitle

\label{firstpage}

\begin{abstract}
Binary star systems containing a neutron star or a black hole with an evolved,
massive star are dynamically perturbed when the latter undergoes a supernova
explosion.
It is possible that the natal kick received by the newly-formed neutron star
in the supernova may place the stellar remnants into a bound, highly eccentric orbit. 
In this case, the two compact objects can tidally interact and 
spiral into one another on a short timescale. 
The interaction with an accretion disc of supernova debris is also considered.
We quantify the likelihood of such events and show that they would be expected
to produce a high-energy transient, possibly a short gamma-ray burst, 
typically within a few days of the supernova. 

\end{abstract}

\begin{keywords}

gamma rays: bursts; binaries: general; X-rays: bursts; 
stars: neutron; black hole: physics; supernovae: general
\end{keywords}

\section{Introduction}\label{sec:intro}

The {\em coalescence} of a neutron star - neutron star (NS-NS) binary 
under the action of gravitational wave emission 
is arguably the leading progenitor 
model for short gamma-ray bursts
\citep[GRBs; e.g.][]{eichler89,meres92,narayan92,rufjan97,rosswog03,oechslin06}.
{\em Collisions} between two neutron stars also do occur,
for instance in regions of large stellar number densities, 
but at a substantially lower rate.
\citet{katz96} suggested  such events might be the central
engines of short GRBs. 

In gravitational wave-driven coalescences 
the extremely large sound velocities in nuclear matter,
$c_s$$\sim$0.4\,c, result in subsonic initial encounters, 
and shocks only form in later phases after rotationally shed material
interacts with itself and with the central object
\citep[][in particular their Fig.~13]{rufjan01,rosswog02a}.
In stark contrast, direct collisions of two neutron stars
approach each other with relative velocities close to free-fall,
$v_{\rm freefall}$\,$\approx$\,$0.64$\,c
$\left(\frac{M}{2.8 {\rm M_{\odot}}}\right)^{1/2}$
$\left(\frac{20\;{\rm km}}{R_1+R_2}\right)^{1/2}$, 
and therefore produce strong shocks passing through 
the nearly unperturbed initial neutron stars.
\citet{rufjan98} found that these shocks seriously pollute the
direct collision vicinity with large amounts of baryonic material, 
which led them to rule out direct neutron star collisions as GRB progenitors. 
However, prospects for producing a GRB may be enhanced for off-centre collisions,
which largely avoid these very strong shocks, and for systems in which one 
of the neutron stars is replaced by a black hole (BH).

Interest in such collisions has been rekindled 
by the observation of prolonged central engine activity 
in some short GRBs. 
Though their nature is still under debate,
features such as X-ray flares \citep{burrows05} 
or long soft tails of emission \citep{barthelmy05,nb06}
suggest continued energy injection. A promising mechanism 
to power the observed late-time activity is the fallback 
accretion of the NS debris \citep{rosswog07,faber06},
and this effect could be enhanced in collisional encounters \citep{lee07}.

Collisions and close encounters between compact objects occur 
predominantly in regions of high stellar density, 
such as the cores of globular clusters \citep[e.g.][]{day76,leonard89,freitag05},
 or the central clusters of massive compact remnants that are expected in
the centres of galaxies \citep[e.g.][]{alexander05}.
This problem has been addressed in a number of previous studies 
\citep{katz96,hansen98}.
In this paper we investigate an alternative possibility of 
compact object collisions induced by supernovae in binary star systems.
We consider binary systems 
initially composed of two massive ($>$8\,\msun) main sequence (MS) stars. 
The primary, i.~e. the more massive star, leaves the MS and explodes 
as a supernova (SN), giving birth to the first compact object, 
either a neutron star or a black hole.
If the binary system does not get disrupted,
we are left with a massive MS star in orbit around a NS (or a BH). 
Since the primary loses most of its envelope mass
prior to the supernova and the secondary is itself a massive star, 
there is a high probability for the binary to survive the explosion
\citep[e.g.][]{vandenh72,pz96}.
 The life of the secondary star will soon come to an end. 
Low mass He stars in wide orbits (P$_{orb}$$>$0.25\,d) 
and those more massive than $\sim$4 \msun~do not go through 
common envelope evolution \citep{dewi03} and the system will undergo the 
second SN explosion. At this point either a) the binary disrupts, or 
b) survives as a NS-NS or a NS-BH binary, or 
c) the newly formed NS is placed in a nearly 
parabolic orbit with a pericenter distance comparable to its tidal radius. 
In the last case we expect the NS to tidally interact with the companion
compact object and rapidly spiral into it. This is the scenario considered here.

Following the second SN, a debris disc could form around the newborn NS 
\citep{zwh08,lin91,cheva89}.
Such discs are thought to be short lived systems 
($\tau_{disc}$$\lesssim$\e{5}\,yrs; \citealt{wck06,exi05}),
therefore they have no influence on NS-NS mergers ($\tau_{merge}$$\sim$Gyr) 
or collisions in globular clusters. 
However, in our case the two compact objects will undergo a close encounter
within days of the explosion and the presence of a fallback disc 
may have a significant effect on the event \citep{popov06}. 

The paper is organized as follows: 
in \S~\ref{sec:prob} we investigate whether collisions induced 
by an asymmetric SN are feasible and how they depend on the binary parameters 
(eccentricity, orbital period, pre-SN mass, SN kick); 
in \S~\ref{sec:delay} we derive the delay time between the
precursor SN and the NSs encounter. The presence
of a debris disc around the newborn NS is addressed in \S~\ref{sec:disc}.
We discuss our results and summarize our conclusions in \S~\ref{end}.

\section{Probability of a collision}\label{sec:prob}

We consider a binary system which already experienced a first SN explosion,
and make two assumptions: 1) the system survived the explosion and
now consists of a compact remnant and the remaining massive star and
2) it is in a circular orbit prior to the second supernova.

\begin{figure}
\centering
\includegraphics[angle=0,scale=0.75]{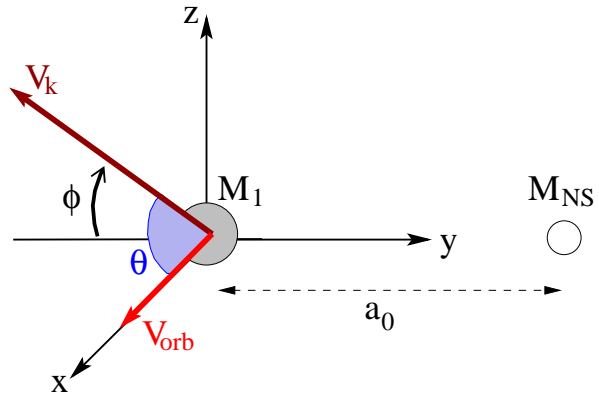}
\caption{Geometry of the system: the reference frame is centered
on the exploding star of mass M$_1$. The NS is at rest, while the
star is moving in a circular orbit, with separation $a_0$ and
relative velocity $V_{orb}$. The pre-SN orbit lies in the $x$-$y$ plane,
perpendicular to the plane of the page.
The angle $\theta$ is defined as the angle between the pre-SN
orbital velocity $V_{orb}$ and the kick velocity $V_{k}$, 
imparted by the SN explosion. The angle $\phi$ is defined as
the angle between the pre-SN orbital plane and the plane 
containing $V_{orb}$ and $V_{k}$.}
\label{geometry}
\end{figure}

The geometry of the system is illustrated in Fig.~\ref{geometry}:
we adopted a reference frame centered on the exploding star of mass M$_1$, 
and where the NS, of mass $M_{NS}$, is at rest. 
The star is moving in a circular orbit, 
with separation $a_0$ and velocity $V_{orb}=(GM_0/a_0)^{1/2}$,
where $G$ is the gravitational constant, 
and $M_0=M_1+M_{NS}$ the total mass of the system. 

The direction of the kick is specified by the two angles
$\theta$ and $\phi$. As shown in Fig.~\ref{geometry}, we defined $\theta$ 
as the angle between the pre-SN
orbital velocity ${\bf V_{orb}}$ and the kick velocity ${\bf V_{k}}$, 
imparted by the SN explosion. The angle $\phi$ is defined as
the angle between the pre-SN orbital plane and the plane 
containing ${\bf V_{orb}}$ and ${\bf V_{k}}$, 
where $\phi=0$ ($\phi=\pi$) if ${\bf V_{k}}$
lies in the pre-SN orbital plane and points outward 
(toward) the NS, and
$\phi=\pi/2$ ($\phi=3\pi/2$) if the plane containing 
${\bf V_{orb}}$ and ${\bf V_{k}}$
is perpendicular to the pre-SN orbital plane and 
${\bf V_{k}}$ has a Cartesian component parallel (anti-parallel)
to the pre-SN angular momentum.
If the kick velocity has no preferential orientation,
then the values of $\cos\theta$ and $\phi$ have an equal probability 
of lying in the range [-1,\,1] and [0,\,2$\pi$], respectively.

To study the effects of the second SN,
we made the standard assumption that the SN mass loss 
and kick are instantaneous 
(i.e. on a timescale shorter than the orbital period), 
and neglected the impact of the expanding SN shell on the compact object 
companion, 
which have little or no effect on the 
binary evolution \citep{kalogera96,fryxell81}.
The specific angular momentum of the system is given by:
\begin{equation}
| {\bf{r}} \times {\bf{V}}| = \left[G M_f~a (1-e^2)\right]^{1/2}
\label{angmom}
\end{equation}
where ${\bf r}$=(0, $a_0$, 0) soon after the SN and 
${\bf V}={\bf V_{orb}}+{\bf V_{k}}$ 
is the resultant velocity of the newborn compact object.
On the right side, $M_f$ is the final mass of the system, 
$a$ and $e$ the post-SN
orbital separation and eccentricity, respectively.
In the following we consider the simplest case of two
identical neutron stars (same mass and radius).

In order to induce a collision the post-explosion orbit
must have a pericenter radius $R_p$=$a(1-e)$ 
comparable to the neutron star radius $R_{NS}$$\approx$10\,km\,$<<$\,$a_0$.
Therefore, only nearly parabolic orbits with eccentricity 
\mbox{$e\gtrsim 1-2R_{NS}/a_0 \approx 1$} may lead to such an encounter.
From Eq.~\ref{angmom} it follows that the kick velocity
must be such to significantly reduce the orbital angular 
momentum of the system. This may happen 
a) if the kick velocity is much higher
than the pre-SN orbital velocity, so that ${\bf{V}}\sim{\bf{V_k}}$,
and directed in the radial direction;
or b) if the kick velocity is comparable in magnitude with
the pre-SN orbital velocity but directed in the opposite
direction, so that ${\bf{V}}\sim{\bf 0}$.

In the former case ($V_k>>V_{orb}$) 
the kick will unbind the binary,
and the newborn NS will collide with the companion
only if the kick direction points exactly toward it. 
The probability of such alignment is small,
$p_{\rm coll}\sim R_{NS}/a_0<$\e{-8}.
Furthermore, such high values of $V_k$ are very
unlikely to occur as they would imply an
unfeasibly large population of isolated neutron stars
for every observed binary pulsar.

A collision between the two compact objects
takes place only if the following constraints on $\theta$
and $\phi$, derived from Eq.~\ref{angmom}, are fulfilled:

\begin{equation}
\sin^2 \phi \lesssim 
\frac{\xi^2-
\left[1+\left(\frac{V_{k}}{V_{orb}}\right) \cos \theta\right]^2}
{\left(\frac{V_{k}}{V_{orb}}\right)^2 \sin^2 \theta} 
\label{sinfi}
\end{equation}

\begin{equation}
- \frac{V_{orb}}{V_k} 
\left(1+ \xi \right) 
\leq \cos\,\theta \leq - \frac{V_{orb}}{V_k} 
\left(1-\xi \right)  
\label{costh}
\end{equation}

where

\begin{equation}
\xi=\left(4 \frac{M_f}{M_0} \frac{R_{NS}}{a_0}\right)^{1/2} << 1 
\label{xi}
\end{equation}

\begin{figure}
\centering
\includegraphics[angle=0,scale=0.5]{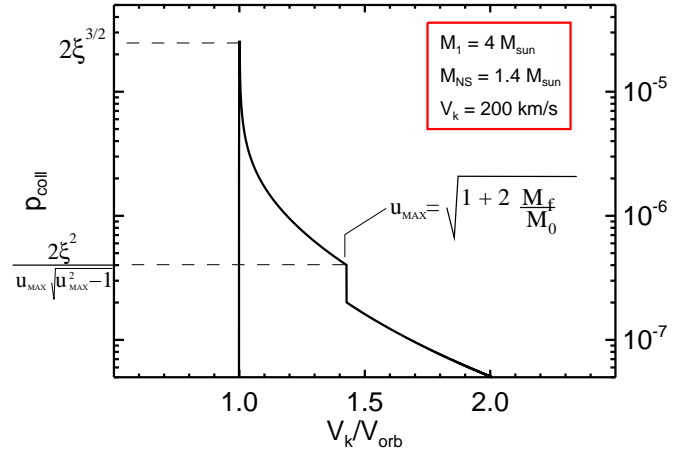}
\caption{Probability of a collision as a function
of the ratio $V_k$/$V_{orb}$ between the kick velocity and 
the pre-SN orbital velocity.
For a given mass loss, the final orbit is bound if
$V_k$/$V_{orb} < u_{MAX}$.
The values of $p_{\rm coll}$ reported on the right $y$-axis
 have been calculated for a kick magnitude $V_k$=200\,\kms, 
a pre-SN mass M$_1$=4\,\msun and a NS mass of 1.4\,\msun.}
\label{prob1}
\end{figure}

It follows immediately from Eq.~\ref{costh} that if $V_k<V_{orb}$
a close encounter is not possible, since $\cos\,\theta \geq -1$,
and that the minimum kick magnitude to induce a collision is
\mbox{$V_k=V_{orb}(1-\xi)\approx V_{orb}$}.
For isotropic kicks, we derived the probability of a collision, $p_{\rm coll}$,
as a function of $V_k$
by numerically integrating the following expression:
\begin{equation}
p_{\rm coll}(V_k) =\frac{1}{\pi} 
\int^{\theta_{min}(V_k)}_{\theta_{MAX}(V_k)} d\theta \sin\theta \int^{\phi_{MAX}(V_k,\theta)}_0 d\phi
\label{prob}
\end{equation}
where the integration limits on $\phi$ and $\theta$ are set by Eq.~\ref{sinfi} 
and \ref{costh}.
The result is shown in Fig.~\ref{prob1}. The values of $p_{\rm coll}$,
quoted on the right $y$-axis, have been calculated for a kick 
magnitude $V_k$=200\,\kms, 
a pre-SN mass M$_1$=4\,\msun and a NS mass M$_{NS}$=1.4\,\msun.

The probability shows a narrow peak at $V_k/V_{orb}$=1
with a maximum value of $p_{\rm coll}$$\sim$2$\xi^{3/2}$, then decreases rapidly.
At $V_k/V_{orb}$=$u_{MAX}$, with $u_{MAX}$=$(1+2M_f/M_0)^{1/2}$,
the probability drops by a factor of two.
In fact, for a given mass loss, highly eccentric post-SN orbits are
still bound only if \mbox{$V_k/V_{orb}<u_{MAX}$}, while higher values of $V_k$
disrupt the system. In the latter case the two compact objects collide 
only if $V_k$ has a Cartesian component directed toward the companion NS,
i. e. only if $\pi/2<\phi<3\pi/2$. 

If the first born compact object is a BH, rather than a NS,
the pericenter radius $R_p$ must be comparable to the NS tidal radius, 
$R_{tid}$$\simeq$$R_{NS}(M_{BH}/M_{NS})^{1/3}$,
in order for the two  objects to interact. 
In addition, the condition $R_p$$>$$R_g$, where $R_g$ is the BH gravitational radius, 
avoids the possibility of the NS being swallowed directly by the BH. 
The latter constraint excludes BHs more massive than $\sim$10\,\msun, 
while for less massive BHs the probability would change by a 
\mbox{factor $\approx(M_{BH}/M_{NS})^{1/4}-(M_{BH}/5~M_{\odot})^{3/4}$}
for a similar value of mass loss.


 \subsection{Natal kicks}\label{sec:kick}

Here we assume that kick velocities follow the birth velocity distribution 
of isolated radio pulsars.
The probability of NS-NS collision is therefore given by the convolution of
the probability $p_{\rm coll}(V_k)$, derived in Eq.~\ref{prob}, 
with the kick velocity distribution $f (V_k)$:
\begin{equation}
P_{\rm coll} = \int_0^\infty p_{\rm coll}(V-V_k) f (V_k) dV_k
\label{maxw}
\end{equation}

\begin{figure}
\centering
\includegraphics[angle=0,scale=0.5]{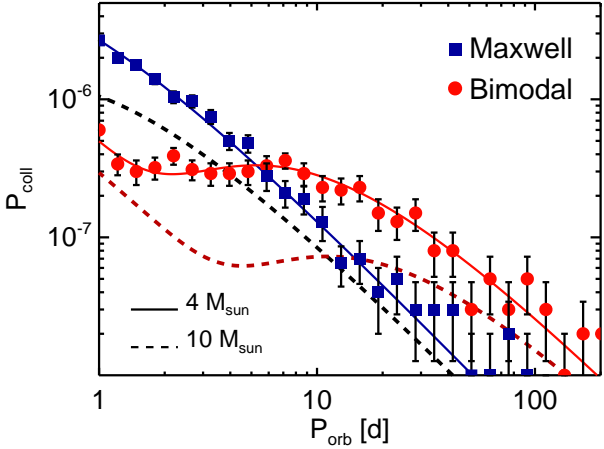}
\caption{Probability of a NS-NS collision as a function
of the pre-SN orbital period $P_{orb}$, calculated 
for a pre-SN He star mass of 4\,\msun~(solid lines)
and 10\,\msun~(dashed lines). The NS mass is 1.4\,\msun.
Symbols report the results of numerical simulations:
we assumed an isotropic distribution of kick directions, 
and ran simulations both for a bimodal distribution (red circles)
and a Maxwellian distribution (blue squares) of kick speeds.
We restricted ourselves to the case in which the binary orbit is
circular prior to the second SN.}
\label{prob2}
\end{figure}

\citet{arzo02} modeled the kick distribution as two Gaussian components 
having characteristic dispersions $\sigma_l$$\sim$90\,km\,s$^{-1}$ 
and $\sigma_h$$\sim$500\,km\,s$^{-1}$.
From a larger sample of 73 young ($<$3\,Myr) pulsars, \citet{hobbs05} 
derived a mean three dimensional birth velocity of $\sim$400\,km\,s$^{-1}$ 
with a characteristic dispersion $\sigma$$\sim$265\,km\,s$^{-1}$. 
The expression in Eq.~\ref{maxw} has been numerically integrated 
both for a bimodal distribution, as in \citet{arzo02} and 
a broad Maxwellian distribution, as in \citet{hobbs05}.
Results are shown in Fig.~\ref{prob2}, which reports the probability of 
a NS-NS collision as a function of the pre-SN orbital period.
Calculations have been performed for a pre-SN mass of 4\,\msun~(solid lines)
and 10\,\msun~(dashed lines).

In order to obtain the distribution of post-SN orbital characteristics
(velocity, separation, eccentricity), we ran Monte-Carlo simulations
of a SN explosion in a circular binary system. We derived the effects 
of a SN on the dynamics of a binary according to the analytic formulation of
\citet{tt98} and the kick direction and magnitude by Monte-Carlo techniques.
We assumed an isotropic distribution of kicks ($P_{\phi}$=1/2$\pi$,
$P_{\theta}$=sin$\theta$/2) and draw the kick magnitude both from the bimodal
and the Maxwell distribution of speeds used in our previous calculations.
For each binary system, we ran \e{8} simulations and estimated the probability 
of a collision as the fraction of simulated  SN explosions 
which lead to an encounter ($R_p\lesssim2R_{NS}$).
In Fig.~\ref{prob2}, the results of the simulations are shown 
as filled squares (Maxwell distribution) and circles (bimodal distribution)
overlaid on the solution of Eq.~\ref{maxw}. The good agreement between
the two confirms our analytic derivation.
 

Throughout this work we assumed that kicks are isotropically distributed
and uncorrelated with the binary properties,
however this is a major source of uncertainty. 
Several physical mechanisms have been proposed
to explain observations of high velocity pulsars 
\citep[and references therein]{lai01,wang06} and, for instance, some
of them invoke a coupling between the direction of the kick
and the stellar spin prior to the explosion \citep{wang07}. 

Fig.~\ref{prob2} shows that the probability of a collision 
event is higher in close binary systems ($P_{orb}$$<$10\,d).
In such close systems the presence of a binary companion can profoundly alter
stellar evolution. 
It seems conceivable that it may somehow affect the SN explosion 
and the kick imparted to the stellar remnant \citep[e.g.][]{pfahl02,pod04}.
Tidal distortion, aspherical stellar winds, 
fast rotation, which in close binaries tends to be aligned 
to the orbital spin, cause asymmetries in the star density 
and velocity distributions. The coupling between the stellar envelope
and its core during the explosion is not understood yet, 
but one might expect  that it can lead to an asymmetric mass ejection, 
hence imparting a recoil velocity which is correlated 
with the orbital motion.

The probability of an encounter would significantly change according 
to the kick geometry.
If natal kicks are preferentially in the direction of the pre-SN orbital angular momentum, 
then the close encounters here described would not happen. 
By contrast, if the preferential direction is along the orbital plane 
in a cone with semi-aperture $\Delta \alpha$, 
the number of collisions would increase by a factor 
$\sim \pi/2 \Delta \alpha$$\lesssim\xi^{1/2}$. 


\begin{figure*}
\centering
\includegraphics[angle=0,scale=0.5]{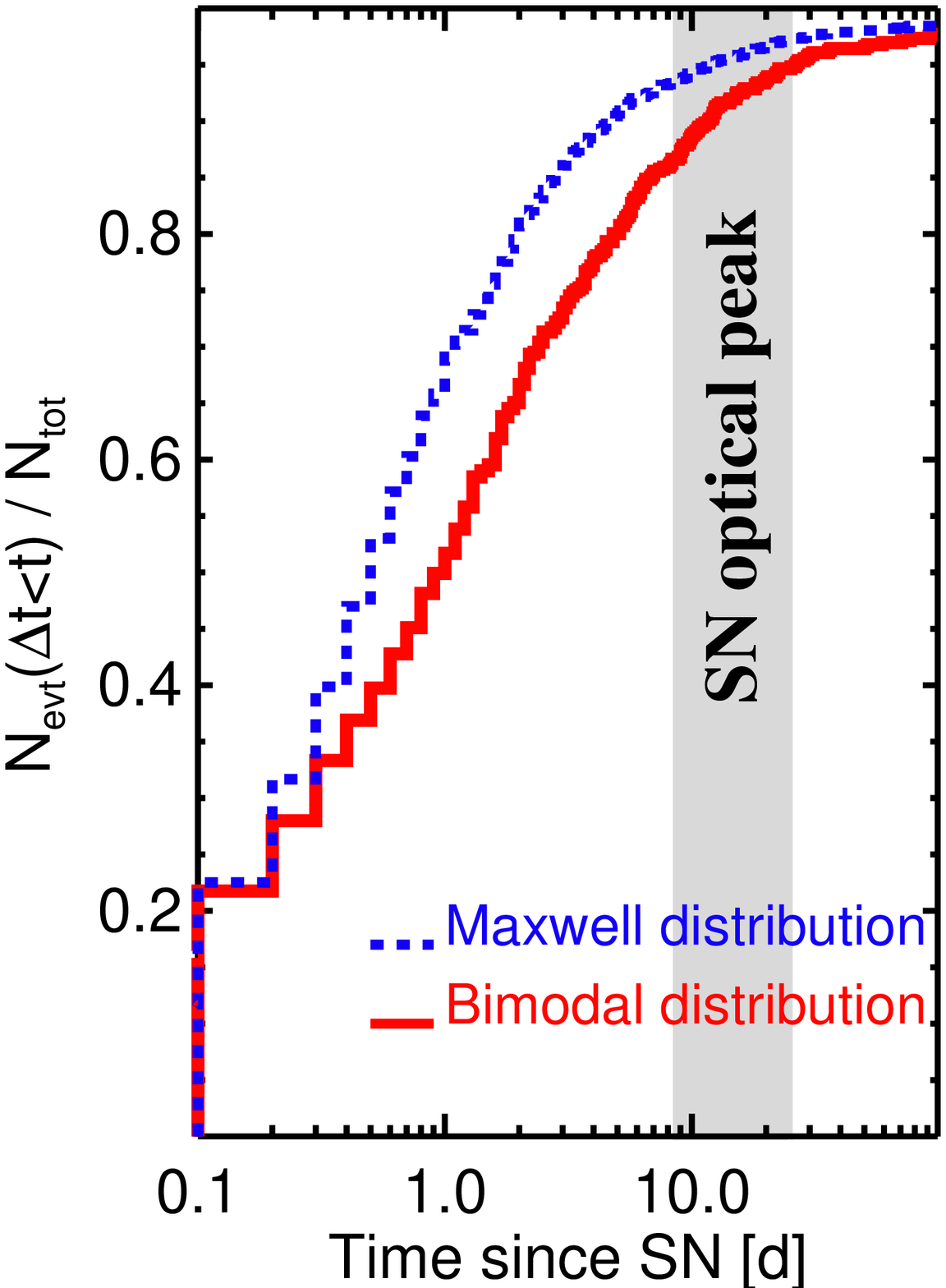}
\hspace{1cm}
\includegraphics[angle=0,scale=0.5]{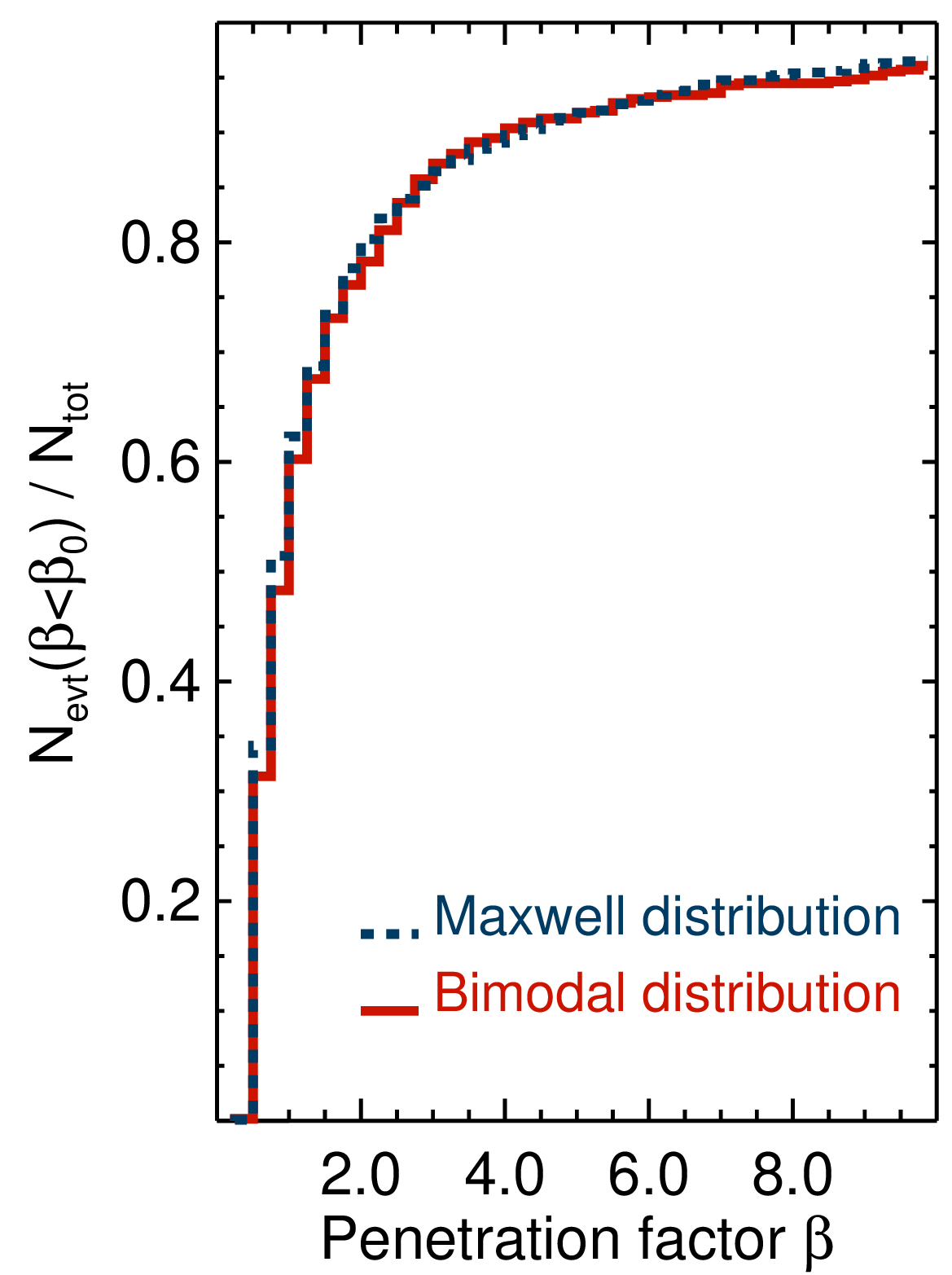}
\caption{$Left~panel$: cumulative distribution of delay times 
since the SN explosion;
the shadowed region indicate the time interval of the SN peak .
$Right~panel$: cumulative distribution of penetration factors, $\beta$,
defined as the ratio between the sum of the NSs radii and the post-SN pericenter
distance, $\beta$=2$R_{NS}$/$R_p$.}
\label{f3}
\end{figure*}


\section{Delay since the supernova}\label{sec:delay}

These collisions are triggered by an asymmetric SN explosion and
will happen soon after it. The time interval $\Delta t$ 
that elapses since the SN can be derived from the known
orbital parameters:

\begin{equation}
\scriptstyle
\Delta t=\sqrt{\frac{a^3}{GM_f}} \left[ \frac{\pi}{2}-\arcsin\left(\frac{1-\frac{a_0}{a}}{e}\right) 
+ \sqrt{-\left(\frac{a_0}{a}\right)^2+2\frac{a_0}{a}-(1-e^2)}\right] 
\end{equation}

where the post-explosion eccentricity $e$ and separation $a$ have been
calculated for each simulated SN following the recipes of \citet{tt98}.
If the newborn NS is shot in the direction opposite to the companion 
($\phi$$\sim$0) but still bound, it will cover almost the entire orbit 
before the encounter; in this case the delay will be 
$\sim 2\pi\sqrt{\frac{a^3}{GM_f}}-\Delta t$.

For hyperbolic orbits ($e\gtrsim1$), we consider only those cases
in which the NS is launched toward the companion ($\phi$$\sim$$\pi$).
The elapsed time since the explosion is:

\begin{equation}
\scriptstyle
\Delta t_{hyp} =\sqrt{\frac{a^3}{GM}} \ln \left\{\left[1+\frac{a_0}{a}+\sqrt{\left(\frac{a_0}{a}\right)^2+2\frac{a_0}{a}-(e^2-1)}\right]\right\}
\end{equation}

Fig.~\ref{f3} shows the cumulative distribution of delays (left panel) 
and penetration factors $\beta$=$2R_{NS}/R_p$ (right panel) 
for the simulations described in \S~\ref{sec:kick}. 
The pre-SN mass of the He star was fixed to 4\,\msun, but
the distributions presented in Fig.~\ref{f3} do not depend on 
this particular choice. In fact the SN mass loss mainly 
affects the normalization of the probability 
curve rather than its shape (see Fig.~\ref{prob2}). 

As shown in Fig.~\ref{f3} (left panel), most of the collisions 
happen a few days after the SN. Though the events are nearly simultaneous, 
the SN will be detectable only a couple of weeks later, 
when it reaches its maximum optical luminosity (shadowed area). 
Only a small fraction ($\lesssim$5\%) of events happen 
at later times ($\Delta t>$30~d). 
Off-axis collisions, which could more easily avoid a
large baryonic contamination, are dominant 
($\sim$80\% with $\beta<2$; Fig.~\ref{f3}, right panel),
while head-on collisions ($\beta>>1$) are rare events ($<$5\%).

\section{The effect of a fallback disc}\label{sec:disc}

In a SN some of the mass is expected to remain bound to the central
object and carry sufficient angular momentum to
settle into a disc \citep{michel88,lin91,woosley93}. 
The presence of such discs around NSs has been invoked to explain 
the peculiar class of anomalous X-ray pulsars (AXPs) and indeed the first 
observational evidence of their existence has been
discovered around the AXP~4U~0142+61 \citep{wck06}. 
Yet it is unclear whether this disc is passive \citep{wck06} or 
affects the properties of the newly formed NS \citep{ertan07}. 
Instead, if a fallback disc persists after a SN in a binary system,
as considered here, it may drastically increase the cross-section
of an interaction event.

Numerical simulations of stellar evolution 
find that the specific angular momentum 
distribution ranges from $j\sim$\e{14}-\e{15}\,cm$^2$\,s$^{-1}$ (in the stellar core) to 
$j\gtrsim$\e{17}\,cm$^2$\,s$^{-1}$ (in the outer layers)
prior to the explosion \citep{heger05}.
The supernova shocks traversing the star
likely induce an extensive mixing between
the stellar layers and an appreciable fraction of 
fallback  material may have an angular momentum 
as high as $j\sim$\e{17}\,cm$^2$\,s$^{-1}$.
By assuming that conservation of angular momentum holds during the collapse, 
this sets the initial size of the fallback disc $R_{NS}$\,$<$\,$R_{d}$\,$<$\,\e{8}\,cm.
For typical values of natal kicks ($<$1000\,\kms; \citealt{chatt05}), the
disc is not left behind, but remains bound to the new born NS. 
The system is therefore composed of a NS (or a BH, formed in the first SN explosion)
and a new born NS (formed in the second SN) surrounded by a fallback disc.

After an initial transient phase the disc spreads to larger radii
as $R_d$\,$\propto$\,$(t/\tau_{\nu})^{3/8}$,
where $\tau_{\nu}\approx$\e{3}-\e{4}\,s is the local viscous timescale 
\citep{lynden74,cannizzo90,menou01},
and reaches a size of $\sim$\e{9}\,cm at 30\,d since the SN.
The chance for the disc to interact with the stellar companion
shortly after the second SN is therefore 
$\propto(R_d/R_{NS})^{3/4}\gtrsim$\,100
times higher than the probability of a direct collision 
derived in \S~\ref{sec:prob}.
The perturbation induced by the passage of the first NS through the disc
extends up to the NS accretion radius 
$r_a=GM_{NS}/v^2_{rel} \approx R_d$, where $v_{rel}$ is the relative velocity 
between the disc and the first NS.
This is comparable with the disc radius itself and therefore
most of the mass in the disc (\e{-6}\,\msun$\lesssim$$M_d$$<$0.1\,\msun;
\citealt{cheva89,lin91,fryer09})
is transferred to the NS on a dynamical timescale of 
$\sim$2\,($R_d$/\e{9} cm)$^{3/2}$\,s.
This episode of accretion may result in a luminous
X-ray outburst a few days/weeks after a SN.
The precise nature of this outburst depends on the details 
of the interaction and the associated gas dynamics but it is 
highly likely to involve accretion rates close to Eddington.

The presence of a debris disc in a binary system
may also lead in some cases to a rapid shrinkage of the orbit,
and speed up the merger of the two objects
\citep{goldreich80,lin86,armitage02}.
For fallback discs of mass $M_d$$<<$0.1\,\msun, 
the characteristic timescale of this process is
$\tau_{shrink} / \tau_{\nu} \gtrsim (M_d+M_{NS})/M_d >> 1$
\citep{lodato09}. The effects of such an interaction 
are not likely to be important since, as pointed out above, most of 
most of the disc mass is transferred to the NS companion
on a much shorter timescale.

\section{Summary and discussion}\label{end}

We showed in \sect{2} that
a supernova may be followed by the collision of two compact objects. 
A close encounter requires the explosion to be asymmetric 
(spherical explosion always led to an expansion of the orbit), 
and happens only for a narrow range of kick magnitudes 
and directions, namely when the kick velocity is comparable to the
pre-SN orbital velocity and opposite in direction.
These events are therefore expected to be rare.

Similarly to mergers, collisions of compact objects end up
forming an accretion disc/black hole system \citep{lee07}, 
as commonly invoked as the central engine powering GRBs.
One of the most important open issue related to (all) the GRB central engines 
is the ``baryonic contamination'' problem \citep[][and references therein]{piran99}: 
How can a violent explosive event involving a few solar masses 
of baryons channel a huge amount of energy into
a region that is almost completely baryon-depleted? 
The popular answer to this question in the context of 
{\em neutron star mergers} has been the centrifugally
evacuated funnel region that forms above the poles of the central remnant
\citep[e.g.][]{davies94,rosswog00,rufjan01},
although it had been realized relatively early on 
that this loophole could possibly be threatened by the ablation
of baryonic material via the extreme neutrino luminosities of $\sim$\,10$^{53}$\,erg/s
\citep{rufjan99,rosswog02b}. 
The first calculations that account for neutrino heating processes 
in a NS merger remnant \citep{dessart09} seriously downsize 
this loophole. For at least as long as
the central, super-massive NS has not collapsed into a black hole, the
remnant drives a very strong baryonic wind ($\dot{\rm M}_w$$\sim 10^{-3}$ M$_\odot$\,s$^{-1}$)  
into exactly the above mentioned funnel region, making the launch of 
a relativistic outflow
impossible. This conclusion may need modification once a black hole
forms.

For the case of {\em neutron star collisions} the situation is even less promising. 
At least head-on collisions ($\beta$$>>$1)
with their violent shocks heavily spoil their surroundings with baryonic material
even if neutrino-driven winds are disregarded \citep{rufjan97}. 
However, we have shown that only a small fraction ($<$5\%) of collisions
are expected to be near-central (cf. Fig.~\ref{f3}, right panel), 
while most of them off-centre.
In this case the outcome may be more favorable to GRBs. 
Off-centre collisions lead to much weaker shocks, 
and a much smoother subsequent coalescence, 
likely dispersing a substantial amount of debris material 
in highly eccentric orbits. This material in turn can produce
late-time X-ray flares simialr to those seen in GRBs \citep{rosswog07,lee07}. 
Collisions between a stellar mass BH and a NS
are even more likely to produce a short GRB since they
avoid two problems from the beginning: there will only be moderate shocks in the disk
formation phase after the disruption and there is never a hot, massive NS as a
major source for the neutrino wind. 
So in the latter cases it may be possible to drive
the ultra-relativistic outflow to power a short-duration GRB.

In the case of NS collisions induced by a SN explosion, 
for a GRB to occur the emerging jet must have a 
minimum power to break out of the remnant layers maintaining 
a high Lorentz factor \citep{macf01,matzner03,zwh04}.
This condition is particularly restrictive when the delay time between the SN
and the collision event is short (t$<<$1~d).
As the supernova remnant (SNR) expands, the requirements
on the jet energy and beaming angle decrease, therefore even low luminosity
and less collimated jets could pass this constraint.
Furthermore, if the SNR radius is larger than the radius for internal shocks
($R_{IS}\approx$\e{12}-\e{13}\,cm), 
the prompt GRB emission can plausibly occur within it,
as the precursor SN has the advantage to clean the 
circumburst environment from baryon pollution \citep{vietri98}, 
thus favoring a relativistic expansion of the GRB jet within the remnant. 
A major problem for longer time delays is whether the GRB emission can
be observed or it is obscured by the dense SNR shell.  
In fact, for a radial velocity of the ejecta v$_{ej}$$\sim$10$^4$\,\kms, 
the SNR radius is $R\sim10^{14}$\,cm after 1 d, 
and the resulting column density is $n_H$$\sim$$10^{28}$(M$_{shell}$/\msun)\,cm$^{-2}$.

Both these problems (the jet break-out and the SNR optical thickness) are not
present if the geometry of the SNR shell is not spherical, but anisotropic
either because of an asymmetric explosion \citep{leonard06,maeda08}
or because subject to hydrodynamic instabilities \citep{cheva92,blondin01}.
The SNR might not interfer with the jet and 
the GRB emission may escape from underdense regions.
In this case, a short GRB (or an X-ray outburst, 
see \sect\ref{sec:disc}) could follow the SN explosion. 
In contrast to long GRBs, connected to the peculiar broad line Type Ic SNe
\citep{wb06}, bursts originated through this channel 
will be associated with any type of core-collapse SN.
The supernova always preceeds the NSs close encounter, but 
according to the distribution of delay times shown in Fig.~\ref{f3}
the high energy transient should be observed first, 
while the SN emission will be visible a few days later. 
For small kick velocities a longer delay between the SN and the
NSs collision is also possible: 
$\sim$10\% of our events happen a couple of weeks after the stellar collapse, 
and a small fraction, $<$5\%, happens more than one month later. 

According to the proposed scenario, double NSs and close encounters
have the same progenitor systems, i.e. a massive binary
system which undergoes two SN explosions.
A fraction of NS-NS binaries will merge within a Hubble timescale,
giving birth to a short GRB \citep{eichler89}. 
By assuming that the main progenitors of short GRBs are binary mergers
(either NS-NS or NS-BH), the observed rate of collisions
$\dot{n}_{coll}^{obs}$ is simply proportional 
to the rate of short GRBs $\dot{n}_{SGRBs}$:
\begin{equation}
\dot{n}_{coll}^{obs} = f_{\theta_c}\dot{n}_{coll} =\frac{n_{coll}}{n_{mergers}}~~f_{\theta_m}~\dot{n}_{SGRBs}
\end{equation}\\
where $f_{\theta_c}$ and $f_{\theta_m}$ are the beaming corrections for bursts 
produced through collisions and mergers, respectively; 
$n_{coll}/n_{mergers}$ is the relative strength of the two channels, 
the former giving rise to prompt collisions and the latter leading to merging NSs.
Following the method described in \citet{kalogera00}
we derived an upper limit of $n_{coll}/n_{mergers}$$\lesssim$\e{-3}.
With a full-sky rate of  $\sim$170 short GRBs per year, 
as derived from the BATSE GRB Catalogue\footnote{http://www.batse.msfc.nasa.gov/batse/grb/catalog/},
and by assuming the same beaming factor of short GRBs originated from double NSs 
($f_{\theta_c}$=$f_{\theta_m}$), 
we estimated for our events an observed rate of  $\lesssim$0.2 yr$^{-1}$.
This number increases by more than a factor of \e{3} when considering 
the interaction with a disc of supernova debris, which may produce
a luminous X-ray outburst.

A wealth of new transient phenomena are to be discovered in the near future,
when the sky will be continously monitored both at high energies (e.g. 
{\it Swift, Fermi, MAXI}) and at optical/NIR wavelenghts (e.g. {\it PAN-STARRS, LSST}). 
Collisions induced by a SN explosion, though rare, might be detected by 
forthcoming surveys.
Such events will be located in young stellar environments, such as
stellar star clusters or star forming galaxies, or
likely at redshift $z\gtrsim1$, when the star formation
rate had its maximum \citep[e.g.][]{madau00,hb06}. 
Though so far short GRBs have been found at moderately low redshifts
$<z>\sim0.5$ and in regions of low/moderate star formation, 
we note that in the growing sample of GRBs with known redshifts
a few bursts, characterized by a rather short intrinsic duration
(e.g. 090426 at $z$=2.6; \citealt{leves09}), may have 
originated from the direct collisions described in this paper.



\section*{Acknowledgments}

We thank A.~R.~King, G.~Lodato and G.~Peres
for valuable comments and suggestions.\\ 
This work has been supported at the University of Leicester by the Science
and Technology Facilities Council, and at INAF by funding from ASI on
grant number I/R/039/04 and by COFIN MIUR grant prot. number
2005025417.  
ET acknowledges the support of the Royal Astronomical Society
during her stay at the University of Leicester.

%

\bibliography{ref-cco}

\begin{thebibliography}{72}
\expandafter\ifx\csname natexlab\endcsname\relax\def\natexlab#1{#1}\fi

\bibitem[{{Alexander}(2005)}]{alexander05}
{Alexander}, T. 2005, \physrep, 419, 65

\bibitem[{{Armitage} \& {Natarajan}(2002)}]{armitage02}
{Armitage}, P.~J. \& {Natarajan}, P. 2002, \apjl, 567, L9

\bibitem[{{Arzoumanian} {et~al.}(2002){Arzoumanian}, {Chernoff}, \&
  {Cordes}}]{arzo02}
{Arzoumanian}, Z., {Chernoff}, D.~F., \& {Cordes}, J.~M. 2002, \apj, 568, 289

\bibitem[{{Barthelmy} {et~al.}(2005){Barthelmy}, {Chincarini}, {Burrows},
  {Gehrels}, {Covino}, {Moretti}, {Romano}, {O'Brien}, {Sarazin},
  {Kouveliotou}, {Goad}, {Vaughan}, {Tagliaferri}, {Zhang}, {Antonelli},
  {Campana}, {Cummings}, {D'Avanzo}, {Davies}, {Giommi}, {Grupe}, {Kaneko},
  {Kennea}, {King}, {Kobayashi}, {Melandri}, {Meszaros}, {Nousek}, {Patel},
  {Sakamoto}, \& {Wijers}}]{barthelmy05}
{Barthelmy}, S.~D., {et~al.} 2005, \nat, 438, 994

\bibitem[{{Blondin} \& {Ellison}(2001)}]{blondin01}
{Blondin}, J.~M. \& {Ellison}, D.~C. 2001, \apj, 560, 244

\bibitem[{{Burrows} {et~al.}(2005){Burrows}, {Romano}, {Falcone}, {Kobayashi},
  {Zhang}, {Moretti}, {O'Brien}, {Goad}, {Campana}, {Page}, {Angelini},
  {Barthelmy}, {Beardmore}, {Capalbi}, {Chincarini}, {Cummings}, {Cusumano},
  {Fox}, {Giommi}, {Hill}, {Kennea}, {Krimm}, {Mangano}, {Marshall},
  {M{\'e}sz{\'a}ros}, {Morris}, {Nousek}, {Osborne}, {Pagani}, {Perri},
  {Tagliaferri}, {Wells}, {Woosley}, \& {Gehrels}}]{burrows05}
{Burrows}, D.~N., {et~al.} 2005, Science, 309, 1833

\bibitem[{{Cannizzo} {et~al.}(1990){Cannizzo}, {Lee}, \&
  {Goodman}}]{cannizzo90}
{Cannizzo}, J.~K., {Lee}, H.~M., \& {Goodman}, J. 1990, \apj, 351, 38

\bibitem[{{Chatterjee} {et~al.}(2005){Chatterjee}, {Vlemmings}, {Brisken},
  {Lazio}, {Cordes}, {Goss}, {Thorsett}, {Fomalont}, {Lyne}, \&
  {Kramer}}]{chatt05}
{Chatterjee}, S., {et~al.} 2005, \apjl, 630, L61

\bibitem[{{Chevalier}(1989)}]{cheva89}
{Chevalier}, R.~A. 1989, \apj, 346, 847

\bibitem[{{Chevalier} {et~al.}(1992){Chevalier}, {Blondin}, \&
  {Emmering}}]{cheva92}
{Chevalier}, R.~A., {Blondin}, J.~M., \& {Emmering}, R.~T. 1992, \apj, 392, 118

\bibitem[{{Davies} {et~al.}(1994){Davies}, {Benz}, {Piran}, \&
  {Thielemann}}]{davies94}
{Davies}, M.~B., {Benz}, W., {Piran}, T., \& {Thielemann}, F.~K. 1994, \apj,
  431, 742

\bibitem[{{Dessart} {et~al.}(2009){Dessart}, {Ott}, {Burrows}, {Rosswog}, \&
  {Livne}}]{dessart09}
{Dessart}, L., {Ott}, C.~D., {Burrows}, A., {Rosswog}, S., \& {Livne}, E. 2009,
  \apj, 690, 1681

\bibitem[{{Dewi} \& {Pols}(2003)}]{dewi03}
{Dewi}, J.~D.~M. \& {Pols}, O.~R. 2003, \mnras, 344, 629

\bibitem[{{Eichler} {et~al.}(1989){Eichler}, {Livio}, {Piran}, \&
  {Schramm}}]{eichler89}
{Eichler}, D., {Livio}, M., {Piran}, T., \& {Schramm}, D.~N. 1989, \nat, 340,
  126

\bibitem[{{Ek{\c s}i} {et~al.}(2005){Ek{\c s}i}, {Hernquist}, \&
  {Narayan}}]{exi05}
{Ek{\c s}i}, K.~Y., {Hernquist}, L., \& {Narayan}, R. 2005, \apjl, 623, L41

\bibitem[{{Ertan} {et~al.}(2007){Ertan}, {Erkut}, {Ek{\c s}i}, \&
  {Alpar}}]{ertan07}
{Ertan}, {\"U}., {Erkut}, M.~H., {Ek{\c s}i}, K.~Y., \& {Alpar}, M.~A. 2007,
  \apj, 657, 441

\bibitem[{{Faber} {et~al.}(2006){Faber}, {Baumgarte}, {Shapiro}, \&
  {Taniguchi}}]{faber06}
{Faber}, J.~A., {Baumgarte}, T.~W., {Shapiro}, S.~L., \& {Taniguchi}, K. 2006,
  \apjl, 641, L93

\bibitem[{{Freitag} \& {Benz}(2005)}]{freitag05}
{Freitag}, M. \& {Benz}, W. 2005, \mnras, 358, 1133

\bibitem[{{Fryer}(2009)}]{fryer09}
{Fryer}, C.~L. 2009, \apj, 699, 409

\bibitem[{{Fryxell} \& {Arnett}(1981)}]{fryxell81}
{Fryxell}, B.~A. \& {Arnett}, W.~D. 1981, \apj, 243, 994

\bibitem[{{Goldreich} \& {Tremaine}(1980)}]{goldreich80}
{Goldreich}, P. \& {Tremaine}, S. 1980, \apj, 241, 425

\bibitem[{{Hansen} \& {Murali}(1998)}]{hansen98}
{Hansen}, B.~M.~S. \& {Murali}, C. 1998, \apjl, 505, L15+

\bibitem[{{Heger} {et~al.}(2005){Heger}, {Woosley}, \& {Spruit}}]{heger05}
{Heger}, A., {Woosley}, S.~E., \& {Spruit}, H.~C. 2005, \apj, 626, 350

\bibitem[{{Hills} \& {Day}(1976)}]{day76}
{Hills}, J.~G. \& {Day}, C.~A. 1976, \aplett, 17, 87

\bibitem[{{Hobbs} {et~al.}(2005){Hobbs}, {Lorimer}, {Lyne}, \&
  {Kramer}}]{hobbs05}
{Hobbs}, G., {Lorimer}, D.~R., {Lyne}, A.~G., \& {Kramer}, M. 2005, \mnras,
  360, 974

\bibitem[{{Hopkins} \& {Beacom}(2006)}]{hb06}
{Hopkins}, A.~M. \& {Beacom}, J.~F. 2006, \apj, 651, 142

\bibitem[{{Kalogera}(1996)}]{kalogera96}
{Kalogera}, V. 1996, \apj, 471, 352

\bibitem[{{Kalogera} \& {Lorimer}(2000)}]{kalogera00}
{Kalogera}, V. \& {Lorimer}, D.~R. 2000, \apj, 530, 890

\bibitem[{{Katz} \& {Canel}(1996)}]{katz96}
{Katz}, J.~I. \& {Canel}, L.~M. 1996, \apj, 471, 915

\bibitem[{{Lai}(2001)}]{lai01}
{Lai}, D. 2001, in Lecture Notes in Physics, Berlin Springer Verlag, Vol. 578,
  Physics of Neutron Star Interiors, ed. D.~{Blaschke}, N.~K. {Glendenning}, \&
  A.~{Sedrakian}, 424--+

\bibitem[{{Lee} \& {Ramirez-Ruiz}(2007)}]{lee07}
{Lee}, W.~H. \& {Ramirez-Ruiz}, E. 2007, New Journal of Physics, 9, 17

\bibitem[{{Leonard} {et~al.}(2006){Leonard}, {Filippenko}, {Ganeshalingam},
  {Serduke}, {Li}, {Swift}, {Gal-Yam}, {Foley}, {Fox}, {Park}, {Hoffman}, \&
  {Wong}}]{leonard06}
{Leonard}, D.~C., {et~al.} 2006, \nat, 440, 505

\bibitem[{{Leonard}(1989)}]{leonard89}
{Leonard}, P.~J.~T. 1989, \aj, 98, 217

\bibitem[{{Levesque} {et~al.}(2009){Levesque}, {Bloom}, {Butler}, {Perley},
  {Cenko}, {Prochaska}, {Kewley}, {Bunker}, {Chen}, {Chornock}, {Filippenko},
  {Glazebrook}, {Lopez}, {Masiero}, {Modjaz}, {Morgan}, \&
  {Poznanski}}]{leves09}
{Levesque}, E.~M., {et~al.} 2009, ArXiv e-prints

\bibitem[{{Lin} \& {Papaloizou}(1986)}]{lin86}
{Lin}, D.~N.~C. \& {Papaloizou}, J. 1986, \apj, 309, 846

\bibitem[{{Lin} {et~al.}(1991){Lin}, {Woosley}, \& {Bodenheimer}}]{lin91}
{Lin}, D.~N.~C., {Woosley}, S.~E., \& {Bodenheimer}, P.~H. 1991, \nat, 353, 827

\bibitem[{{Lodato} {et~al.}(2009){Lodato}, {Nayakshin}, {King}, \&
  {Pringle}}]{lodato09}
{Lodato}, G., {Nayakshin}, S., {King}, A.~R., \& {Pringle}, J.~E. 2009, ArXiv
  e-prints

\bibitem[{{Lynden-Bell} \& {Pringle}(1974)}]{lynden74}
{Lynden-Bell}, D. \& {Pringle}, J.~E. 1974, \mnras, 168, 603

\bibitem[{{MacFadyen} {et~al.}(2001){MacFadyen}, {Woosley}, \&
  {Heger}}]{macf01}
{MacFadyen}, A.~I., {Woosley}, S.~E., \& {Heger}, A. 2001, \apj, 550, 410

\bibitem[{{Madau} \& {Pozzetti}(2000)}]{madau00}
{Madau}, P. \& {Pozzetti}, L. 2000, \mnras, 312, L9

\bibitem[{{Maeda} {et~al.}(2008){Maeda}, {Kawabata}, {Mazzali}, {Tanaka},
  {Valenti}, {Nomoto}, {Hattori}, {Deng}, {Pian}, {Taubenberger}, {Iye},
  {Matheson}, {Filippenko}, {Aoki}, {Kosugi}, {Ohyama}, {Sasaki}, \&
  {Takata}}]{maeda08}
{Maeda}, K., {et~al.} 2008, Science, 319, 1220

\bibitem[{{Matzner}(2003)}]{matzner03}
{Matzner}, C.~D. 2003, \mnras, 345, 575

\bibitem[{{Menou} {et~al.}(2001){Menou}, {Perna}, \& {Hernquist}}]{menou01}
{Menou}, K., {Perna}, R., \& {Hernquist}, L. 2001, \apj, 559, 1032

\bibitem[{{Meszaros} \& {Rees}(1992)}]{meres92}
{Meszaros}, P. \& {Rees}, M.~J. 1992, \apj, 397, 570

\bibitem[{{Michel}(1988)}]{michel88}
{Michel}, F.~C. 1988, \nat, 333, 644

\bibitem[{{Narayan} {et~al.}(1992){Narayan}, {Paczynski}, \&
  {Piran}}]{narayan92}
{Narayan}, R., {Paczynski}, B., \& {Piran}, T. 1992, \apjl, 395, L83

\bibitem[{{Norris} \& {Bonnell}(2006)}]{nb06}
{Norris}, J.~P. \& {Bonnell}, J.~T. 2006, \apj, 643, 266

\bibitem[{{Oechslin} \& {Janka}(2006)}]{oechslin06}
{Oechslin}, R. \& {Janka}, H.-T. 2006, \mnras, 368, 1489

\bibitem[{{Pfahl} {et~al.}(2002){Pfahl}, {Rappaport}, {Podsiadlowski}, \&
  {Spruit}}]{pfahl02}
{Pfahl}, E., {Rappaport}, S., {Podsiadlowski}, P., \& {Spruit}, H. 2002, \apj,
  574, 364

\bibitem[{{Piran}(1999)}]{piran99}
{Piran}, T. 1999, \physrep, 314, 575

\bibitem[{{Podsiadlowski} {et~al.}(2004){Podsiadlowski}, {Langer},
  {Poelarends}, {Rappaport}, {Heger}, \& {Pfahl}}]{pod04}
{Podsiadlowski}, P., {Langer}, N., {Poelarends}, A.~J.~T., {Rappaport}, S.,
  {Heger}, A., \& {Pfahl}, E. 2004, \apj, 612, 1044

\bibitem[{{Popov}(2006)}]{popov06}
{Popov}, S.~B. 2006, arXiv:astro-ph/0610593

\bibitem[{{Portegies Zwart} \& {Verbunt}(1996)}]{pz96}
{Portegies Zwart}, S.~F. \& {Verbunt}, F. 1996, \aap, 309, 179

\bibitem[{{Rosswog}(2007)}]{rosswog07}
{Rosswog}, S. 2007, \mnras, 376, L48

\bibitem[{{Rosswog} \& {Davies}(2002)}]{rosswog02a}
{Rosswog}, S. \& {Davies}, M.~B. 2002, \mnras, 334, 481

\bibitem[{{Rosswog} {et~al.}(2000){Rosswog}, {Davies}, {Thielemann}, \&
  {Piran}}]{rosswog00}
{Rosswog}, S., {Davies}, M.~B., {Thielemann}, F.-K., \& {Piran}, T. 2000, \aap,
  360, 171

\bibitem[{{Rosswog} \& {Ramirez-Ruiz}(2002)}]{rosswog02b}
{Rosswog}, S. \& {Ramirez-Ruiz}, E. 2002, \mnras, 336, L7

\bibitem[{{Rosswog} {et~al.}(2003){Rosswog}, {Ramirez-Ruiz}, \&
  {Davies}}]{rosswog03}
{Rosswog}, S., {Ramirez-Ruiz}, E., \& {Davies}, M.~B. 2003, \mnras, 345, 1077

\bibitem[{{Ruffert} \& {Janka}(1997)}]{rufjan97}
{Ruffert}, M. \& {Janka}, H.-T. 1997, in Reviews in Modern Astronomy, Vol.~10,
  Reviews in Modern Astronomy, ed. R.~E. {Schielicke}, 201--218

\bibitem[{{Ruffert} \& {Janka}(1998)}]{rufjan98}
{Ruffert}, M. \& {Janka}, H.-T. 1998, \aap, 338, 535

\bibitem[{{Ruffert} \& {Janka}(1999)}]{rufjan99}
{Ruffert}, M. \& {Janka}, H.-T. 1999, \aap, 344, 573

\bibitem[{{Ruffert} \& {Janka}(2001)}]{rufjan01}
{Ruffert}, M. \& {Janka}, H.-T. 2001, \aap, 380, 544

\bibitem[{{Tauris} \& {Takens}(1998)}]{tt98}
{Tauris}, T.~M. \& {Takens}, R.~J. 1998, \aap, 330, 1047

\bibitem[{{van den Heuvel} \& {Heise}(1972)}]{vandenh72}
{van den Heuvel}, E.~P.~J. \& {Heise}, J. 1972, \nat, 239, 67

\bibitem[{{Vietri} \& {Stella}(1998)}]{vietri98}
{Vietri}, M. \& {Stella}, L. 1998, \apjl, 507, L45

\bibitem[{{Wang} {et~al.}(2006{\natexlab{a}}){Wang}, {Lai}, \& {Han}}]{wang06}
{Wang}, C., {Lai}, D., \& {Han}, J.~L. 2006{\natexlab{a}}, \apj, 639, 1007

\bibitem[{{Wang} {et~al.}(2007){Wang}, {Lai}, \& {Han}}]{wang07}
{Wang}, C., {Lai}, D., \& {Han}, J.~L. 2007, \apj, 656, 399

\bibitem[{{Wang} {et~al.}(2006{\natexlab{b}}){Wang}, {Chakrabarty}, \&
  {Kaplan}}]{wck06}
{Wang}, Z., {Chakrabarty}, D., \& {Kaplan}, D.~L. 2006{\natexlab{b}}, \nat,
  440, 772

\bibitem[{{Woosley}(1993)}]{woosley93}
{Woosley}, S.~E. 1993, \apj, 405, 273

\bibitem[{{Woosley} \& {Bloom}(2006)}]{wb06}
{Woosley}, S.~E. \& {Bloom}, J.~S. 2006, \araa, 44, 507

\bibitem[{{Zhang} {et~al.}(2004){Zhang}, {Woosley}, \& {Heger}}]{zwh04}
{Zhang}, W., {Woosley}, S.~E., \& {Heger}, A. 2004, \apj, 608, 365

\bibitem[{{Zhang} {et~al.}(2008){Zhang}, {Woosley}, \& {Heger}}]{zwh08}
{Zhang}, W., {Woosley}, S.~E., \& {Heger}, A. 2008, \apj, 679, 639

\end{thebibliography}

\end{document}